\documentclass[useAMS,usenatbib,usegraphicx]{mn2e}
\include{epsf}

\newcommand\aj{{AJ}}%
\newcommand\araa{{ARA\&A}}%
\newcommand\apj{{ApJ}}%
\newcommand\apjl{{ApJ}}%
\newcommand\apjs{{ApJS}}%
\newcommand\aap{{A\&A}}%
\newcommand\aaps{{A\&AS}}%
\newcommand\mnras{{MNRAS}}%
\newcommand\pasp{{PASP}}%
\newcommand\pasj{{PASJ}}%
\newcommand\nat{{Nature}}%
\title[Bubble-Induced Star Formation in dIrrs]
{Numerical Simulations of Bubble-Induced Star Formation in Dwarf Irregular Galaxies with a Novel Stellar Feedback Scheme}
\author[Kawata et~al.]
 {\parbox{\textwidth}{Daisuke~Kawata$^{1}$\thanks{E-mail: d.kawata@ucl.ac.uk},
Brad~K.~Gibson$^{2}$, David~J.~Barnes$^{1}$, Robert~J.J.~Grand$^{1}$,
 Awat Rahimi$^{3}$}\vspace{0.5cm}
\\
$^{1}$ Mullard Space Science Laboratory, University College London,
Holmbury St. Mary, Dorking, Surrey, RH5 6NT, UK
\\
$^{2}$ Jeremiah Horrocks Institute, 
University of Central Lancashire, Preston, PR1~2HE, UK
\\
$^{3}$ National Astronomical Observatory, Chinese Academy of Science, 20A Datun Road, Beijing, China
}
\date{Accepted .
      Received ;
      in original form }

\pagerange{\pageref{firstpage}--\pageref{lastpage}}
\pubyear{2013}

\begin{document}

\maketitle

\label{firstpage}

\begin{abstract}
To study the star formation and feedback mechanism, we simulate the evolution of an isolated dwarf irregular galaxy (dIrr) in a fixed dark matter halo, similar in size to WLM, using a new stellar feedback scheme. We use the new version of our original N-body/smoothed particle chemodynamics code, {\tt GCD+}, which adopts improved hydrodynamics, metal diffusion between the gas particles and new modelling of star formation and stellar wind and supernovae (SNe) feedback. Comparing the simulations with and without stellar feedback effects, we demonstrate that the collisions of bubbles produced by strong feedback can induce star formation in a more widely spread area. We also demonstrate that the metallicity in star forming regions is kept low due to the mixing of the metal-rich bubbles and the metal-poor inter-stellar medium. Our simulations also suggest that the bubble-induced star formation leads to many counter-rotating stars. 
The bubble-induced star formation could be a dominant mechanism to maintain star formation in dIrrs, which is different from larger spiral galaxies where the non-axisymmetric structures, such as spiral arms, are a main driver of star formation.
\end{abstract}


\section{Introduction}
\label{intro-sec}

 Dwarf Irregular galaxies (dIrrs) are powerful laboratories for studying the impact of stellar feedback upon star formation and gas dynamics, because their effects are likely to be more dramatic in a low-mass system \citep[e.g.][]{mf99}. They  (in this paper, we focus on a system whose stellar mass is less than $\sim10^8$ M$_{\odot}$) are generally a gas rich system and sustain star formation for a long time. Some of the dIrrs, e.g. Wolf-Lundmark-Melotte \citep[WLM,][]{mw09,pm26}, are close enough for observers to resolve stellar populations in the galaxies \citep[e.g.][]{mz97,ad00,rmgza00,jsgpw07,behmo12}, and detailed structures of HI gas \citep[e.g][]{hsm81,bdb04,jscc04,kwhn07,hzbwk11} are observed. WLM is often considered to be the prime target for studying the intrinsic properties of a star forming low-mass systems, because it is well isolated from the rest of the galaxies in the Local Group, and it is unlikely that WLM has experienced dynamical interaction with the other galaxies in the recent past \citep[e.g.][]{lvbbc12}. Photometric studies \citep[e.g.][]{mz97,rmgza00,jsgpw07}
 reveal that WLM harbours old stellar populations. \citet{ad00} suggested that WLM created about half of its stars in an initial strong star formation 10-12 Gyr ago. After that, the galaxy maintains a low level of the star formation rate (SFR), $\sim1$-$2\times 10^{-4}$ M$_{\odot}$ yr$^{-1}$ on average with a higher SFR in the last few Gyr. \citet{hel10} suggested a current SFR  of $6\times10^{-3}$ M$_{\odot}$ yr$^{-1}$ from {\it Galaxy Evolution Explorer (GALEX)} imaging data. The metallicity of stars has been measured with spectroscopic observations of individual stars and HII regions. The estimated abundances for the young supergiants show a range of [Z] \footnote{$[{\rm Z}]=\log ({\rm Z}/{\rm Z}_{\odot})$} $\sim-0.5$ dex to $-1$ dex \citep{vtksc03,ukbpg08}, and the mean value of [Z]$=-0.87\pm0.06$ dex is consistent with the measured abundances for HII regions \citep{stm89,hm95,lsv05}. Recently CO has been detected in WLM \citep{erhvb13}. However, the CO abundance is much lower than what is expected from the current SFR and the empirical relation of the Milky Way data. The number of detailed observations are increasing recently for WLM, and even more detailed observations are being produced for other nearby dIrrs \citep[e.g.][]{he06,slmpr06,hel10,sdwso13}. These observations provide strong constraints on the chemical evolution models of dwarf galaxies \citep[e.g.][]{fggl06,bg07}. 
  
  Dwarf galaxies are a good target for numerical simulations which aim to model the inter-stellar medium (ISM), star formation and stellar feedback in a galaxy. As dwarf galaxies are physically small, the physical resolution is higher if the simulations use a similar number of resolution elements to simulations of larger galaxies, like the Milky Way. \citet{mytn97} simulated the formation of a dwarf galaxy as a monolithic collapse of the gas within a dark matter halo. They demonstrated that due to the small gravitational potential of dwarf galaxies, the super-shells created by stellar feedback owing to the star formation in the central region collides with the infalling gas and subsequently forms stars, which can generate the positive colour gradient, similar to some of the observed dwarf ellipticals \citep[e.g.][]{vvls88,tnccj10}. 
 Using cosmological simulations \citet{mrng05} demonstrated that the small galaxies formed at high redshifts can explain the global properties, such as the luminosity, velocity dispersion and metal abundances, of the dwarf galaxies observed in the Local Group \citep[see also][]{kacg06}. 
 
 
 Strong feedback effects on dwarf galaxies are studied in various kinds of numerical simulations \citep[e.g.][]{ccgl01,kwb07,sdqkw07,sdqgk09,rj12,bst13,pgbcs12,sbjsm13,tpdr13,sdkcv13}, and strong feedback is considered to be a key element in building a more realistic dwarf galaxy formation model \citep[e.g.][]{gbm10,sgsjw11}. \citet{ccgl01}, \citet{sdqgk09} and \citet{pgcbm11} demonstrate that because of the lower gravitational potential of dwarf galaxies, stellar feedback creates outflows of the gas and stops star formation. However, if the outflow is not strong enough to escape from the system, the gas will fall back again, and restart star formation. As a result, dwarf galaxies can maintain the intermittent star formation activities, which is called "breathing" star formation activity in \citet{sdqkw07}. Using a monolithic collapse simulation, \citet{kwb07} and \citet{tpdr13} showed that strong stellar feedback makes dwarf disc galaxies thicker and puffier and leads to a more velocity dispersion dominated system with respect to the rotation velocity. 
  
    We have recently updated our original N-body/smoothed particle chemodynamics code, {\tt GCD+} \citep{kg03a,bkw12,rk12,kogbc13}. \citet{kogbc13} demonstrated that the new version of {\tt GCD+} captures hydrodynamics much better than conventional
Smoothed Particle Hydrodynamics (SPH) codes used in the above-mentioned prior studies, especially for shocks induced by point-like explosions, such as a supernova (SN). We also adopt radiative cooling and heating rates, taking into account the metallicity of the gas and UV background radiation, which are ignored or simplified in some of the previous works. This improvement of the code enables us to follow smaller-scale physics more accurately, and justifies the use of higher resolution particles. We carry out high resolution simulations of the evolution of a dIrr, similar in size to WLM, and study how stellar feedback affects the ISM and star formation in small galaxies. DIrr simulations with similar physical resolution were reported in \citet{hqm11,hqm12b,hqm12a}, who focused on the difference between galaxies of different masses. We focus on the detailed picture of how star forming regions affect one another, how dIrrs sustain the low level of star formation and how they maintain a low metallicity.
Section~\ref{meth-sec} describes briefly the new version of {\tt GCD+} and numerical models of our dIrr simulation.
Section~\ref{res-sec} presents the results.
A summary of this study is presented in Section~\ref{sum-sec}.

\section{Method and Models}
\label{meth-sec}

\subsection{New version of GCD+}
\label{code-sec}

{\tt GCD+} is a parallel three-dimensional N-body/SPH \citep{ll77,gm77} code which can be used in studies of galaxy formation and evolution in both a cosmological and isolated setting \citep{kg03a,bkw12,kogbc13}.
{\tt GCD+} incorporates self-gravity, hydrodynamics, radiative cooling, star formation, SNe feedback, and metal enrichment. The new version of the code includes metal diffusion patterned after the formalism described by \citet{ggb09}.
We have implemented a modern scheme of SPH suggested by \citet{rp07}, including their artificial viscosity switch \citep{jm97} and artificial thermal conductivity to resolve
the Kelvin-Helmholtz instability \citep[see also][]{rh12,sm13}. Following \citet{sh02}, we integrate the entropy equation instead of the energy equation. As suggested by \citet{sm09}, we have added the individual time step limiter, which is crucial for correctly resolving the expansion bubbles induced by SNe feedback \citep[see also][]{mbg10,dd12}. We also implement the FAST scheme \citep{sm10} which allows the use of different timesteps for integrating hydrodynamics and gravity. The code also includes adaptive softening for both the stars and gas \citep{pm07}, although in this current study we only implement it for the gas particles. Our recent update and performance in various test problems are presented in \citet{bkw12} and \citet{kogbc13}.

 Radiative cooling and heating are calculated with CLOUDY \citep[v08.00:][]{fkv98} following \citet{rk08}. We tabulate cooling and heating rates and the mean molecular weight as a function of redshift, metallicity, density and temperature adopting the 2005 version of the \citet{hm96} UV background radiation.

We also add a thermal energy floor following \citet{rk08} and \citet{hqm11} to keep the Jeans mass higher than the resolution and avoid numerical instabilities \citep[see also][]{bb97}. Following \citet{hqm11}, we implement the pressure floor,
\begin{equation}
 P_{\rm Jeans}=1.2 N_{\rm Jeans}^{2/3} \gamma^{-1} G h^2 \rho^2,
\label{peffeq}
\end{equation}
where $\gamma=5/3$, and $h$ and $\rho$ are the smoothing length and density of the gas particle.  Following \citet{hqm11}, we set $N_{\rm Jeans}=10$, and this pressure floor is adopted only in the Euler equation, i.e. the temperature of the gas is allowed to cool following the radiative cooling.

We allow gas particle to become a star particle if the pressure of the gas particle is smaller than their $P_{\rm Jeans}$ and the density becomes greater than the star formation threshold density, $n_{\rm H,th}=1000$ cm$^{-3}$ \citep{hkm13}, following the Schmidt law as described in \citet{kg03a}:
\begin{equation}
 \frac{d \rho_*}{dt} = -\frac{d \rho_{\rm g}}{dt}
 = \frac{C_* \rho_{\rm g}}{t_{\rm g}},
\label{sfreq}
\end{equation}
where $C_*$ is a dimensionless SFR parameter and $t_{\rm g}$ is the dynamical time. We set $C_*=1$, which is rather large, compared to $C_*\sim0.02$ often used in these kind of simulations, including our previous work \citep{rk12}. However, note that the pressure of the gas particle becomes less than $P_{\rm Jeans}$ around $n_{\rm H}=10$ cm$^{-3}$. This is our resolution limit, and we set the gravitational softening limit corresponding to this density (see below). 
Therefore, our threshold density for star formation in this paper,  $n_{\rm H,th}=1000$ cm$^{-3}$, is much higher than our resolution limit. As a result, the gas particles experience some delay from the time they hit the resolution limit to the time when the density becomes higher than the star formation threshold. This delay is effectively similar to setting a smaller $C_*$ value and lower star formation threshold density (see also Appendix \ref{lrm-sec}). We think that this is a part of the reason why high density threshold of star formation leads to a resultant SFR independent of the parameter $C_*$ \citep{sdk08,hkm13}\footnote{Note that the pressure floor in equation (\ref{peffeq}) enables us to "resolve" the Jeans mass at a higher density than the resolution limit in our definition, i.e. where the pressure of the gas particle becomes less than $P_{\rm Jeans}$, which helps to justify the high $n_{\rm H,th}$ we adopted \citep{sdv08,rk08}. \citet{sdv08} used a softening length much smaller than the resolution limit, and let the gas clouds collapse in the dynamical time-scale till the density becomes high enough for the softening length to be effective. On the other hand, we set the softening length limit close to the resolution limit, because the smaller softening length demands smaller time steps, and it is too expensive for our high-resolution simulation. As a result, the gas collapse is slowed down, once their density reaches the resolution limit, which provides a similar effect to applying a small $C_*$, because our $n_{\rm H,th}$ is higher than the resolution limit. In other words, the softening length limit is one of the parameters to control SFR.}.
We prefer $C_*=1$ and high $n_{\rm H,th}$ over low $C_*$ and low $n_{\rm H,th}$, because $C_*=1$ allows the gas particles
to turn into the star particles soon after they reach $n_{\rm H,th}$, while low $C_*$ allows the gas particles to survive until they 
reach quite high density, which slows down the simulation significantly. This choice may depend on the resolution and the kind of 
radiative cooling applied.

We assume that stars are distributed according to the \citet{es55} initial mass function (IMF). {\tt GCD+} takes into account chemical enrichment by both Type II SNe \citep[SNe II,][]{ww95} and Type Ia SNe \citep[SNe Ia,][]{ibn99,ktn00} and mass-loss from intermediate-mass stars \citep{vdhg97}, and follows the chemical enrichment history of both the stellar and gas components of the system.  

 \subsection{Stellar Feedback}
 
 The new version of {\tt GCD+} uses a new scheme for star formation and feedback. We keep the mass of the baryon (gas and star) particles equal, unlike \citet{kg03b} or the majority of  SPH simulations which include star formation.
  The basic strategy for our stellar feedback was inspired by \citet{lpc02} and \citet{msd08}. However, a slightly different implementation is adopted. First, every star particle formed in the simulation is randomly assigned a mass group ID ranging from 1 to 61 (although 61 is chosen arbitrarily, it is a compromised selection to sample the stellar mass range and be similar to our resolution, i.e. number of neighbour particles). Our new stellar feedback scheme models the mass, energy and metal feedback using this ID. A star cluster following the assumed IMF and the adopted remnant masses will undergo the mass-loss of about 30 \% of their original stellar mass after the Hubble time. Therefore, star particles whose ID is between 1 and 19 are used to describe the mass-loss, energy feedback and metal enrichment.  Particles with smaller IDs are responsible for feedback from higher mass stars. 
In our feedback scheme 61 particles describe a whole mass range of stars in a star cluster, which corresponds to
6,100 M$_{\odot}$ for the simulation in this paper, where the particle mass is $m_p=100$ M$_{\odot}$.
  
Depending on their ID and age, a star particle becomes a feedback particle for some period as described below, and then becomes a normal gas particle. In other words, mass-loss from stars is modelled by a star particle becoming a normal gas particle, and hence the particle mass of gas and star particle is always equal.

The pressure from the feedback particles affects the neighbouring gas particles through the Euler equation. However,
the feedback particles do not feel any reaction from the neighbour particles, and follow the collisionless N-body equation
for simplicity  \citep[see also][]{pvi04}. 
The thermal energy of the feedback particles is calculated with the SPH scheme with radiative cooling and additional heating from
stellar wind, SNe II and SNe Ia.  Eight out of the 19 feedback particles spend no time being stars and immediately become
SNe II feedback particles (Sec. \ref{sneiip-sec}), which are responsible to describe stellar feedback 
due to stellar wind and SNe II, and no radiative cooling is applied until their age becomes older than the lifetime of 8 M$_{\odot}$  
or their cooling time becomes longer than dynamical time.

The metals produced by SNe II, SNe Ia and intermediate mass stars are also distributed from the feedback particles 
to neighbouring gas particles through the metal diffusion scheme of \citet{ggb09}, i.e. the metal diffusion applied to
the feedback particles is the same as the other normal gas particles. Once the age of the feedback particles 
becomes old enough, the feedback particle becomes a normal gas particle.
Because gas density, temperature and metal abundances for feedback particles are calculated through the SPH scheme,
including metal diffusion, the new gas particles inherit these properties from the feedback particles.

\begin{figure}
\centering
\includegraphics[width=\hsize]{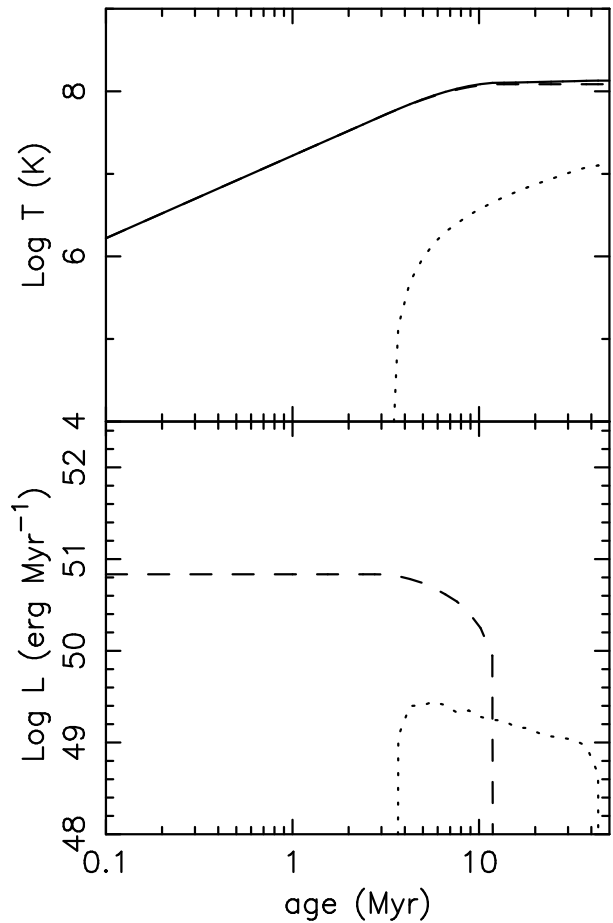}
\caption{
Upper Panel: Temperature evolution due to stellar energy feedback of SNe II particle with solar metallicity as a function of the age (solid line). The contribution from stellar wind and SNe II are also shown with dashed and dotted lines respectively.
Lower Panel: The energy deposition rate (erg~Myr$^{-1}$)
from stellar wind (dashed line) and SNe II (dotted line) to a SNe~II particle of 100 M$_{\odot}$ as a function of the age. 
Stellar wind energy of $L_{\rm SW}=10^{37}$ erg~s$^{-1}$ per massive star ($>15$ M$_{\odot}$)
and SNe~II energy of $E_{\rm SN}=10^{50}$ erg per SN are assumed like our fiducial model.
}
\label{tempt-fig}
\end{figure}

\subsubsection{SNe II Particles}  
\label{sneiip-sec}
  
We can calculate that about 13 \% of the mass is ejected by SNe II from a star cluster following the assumed IMF and the remnant masses shown in \citet{ww95}. 
Therefore, each star particle in the ID range between 1 and 8 is set to be a feedback particle, "SNe~II particle".
SNe~II particles are responsible for the energy and metals produced by stellar winds and SNe II.
SNe~II particles are treated as one kind of particle. In other words, there is no difference between star particles whose ID is less than 
or equal to 8.

 A star particle whose ID is between 1 and 8 becomes a SNe II particle at the same time step at which they form.
This allows all the SNe II particles to be effective immediately and maintain their effect for a long time.
For example, SNe II particles with solar metallicity are responsible for mass-loss and feedback from 100 to 7 M$_{\odot}$ stars. 
Once the age of the SNe II particle becomes older than the lifetime of 7 M$_{\odot}$ star, the SNe II particle becomes a gas particle. 

We assume each SN produces thermal energy $E_{\rm SN}$ (erg). We also assume that stellar wind from each massive star ($>15$ M$_{\odot}$) produces thermal energy $L_{\rm SW}$ (erg s$^{-1}$)\footnote{Feedback due to massive stars is often divided to several mechanisms, such as radiation pressure, photoionization heating, and stellar wind \citep[e.g.][]{hqm12a}. However, we avoid introducing more complicated sub-grid models or
justifications, because sub-grid models are often implementation dependent and associated with more parameters which affect the system
in complex ways.  Rather, we simply call the overall feedback effects (in our resolution and implementation) from massive stars "stellar wind", and use the parameter, $L_{\rm SW}$, to control energy feedback before SNe II kick in.} and add this to the SNe II particles. 
Because there are 8 SNe~II particles out of 61 star particles, each SNe II particle receives 1/8-th of the energy and metals produced by a star cluster of 61 star particles, depending on the age and initial metallicity.
Consequently SNe~II particles have higher temperature and become metal-rich.
Fig.~\ref{tempt-fig} shows the evolution of temperature and the energy deposition rate to
a SNe~II particle with solar metallicity as a function of the age, assuming 
$L_{\rm SW}=10^{37}$ erg~s$^{-1}$ per massive star \citep[e.g.][]{wmcsm77,mjs80,bg94,om94}
 and $E_{\rm SN}=10^{50}$ erg per SN \citep[e.g.][]{tgjs98} which are used in our fiducial model.

The pressure from SNe II particles affects the neighbour particles through the Euler equation, 
while SNe II particles follow the collisionless N-body equation.
We calculate thermal energy for the SNe II particles following the SPH scheme, and radiative cooling is turned off.
Therefore, although the temperature of the SNe II particle can increase, e.g. following Fig.~\ref{tempt-fig},
SNe II particles is subject to adiabatic expansion and compression,
and their temperature evolution does not follow Fig.~\ref{tempt-fig} completely.

We turned off radiative cooling until their age becomes older than the lifetime of 8 M$_{\odot}$  
or their cooling time becomes longer than dynamical time.
This is similar in spirit to the blastwave feedback \citep{myn99,tc00,bkgf04a,ssk06}.
Note that we do not adopt the kinetic feedback scheme \citep[e.g.][]{nw93,kg03a,sh03}. 

We do not distribute the feedback energy to the neighbour particles, only the SNe II particle itself 
receives the thermal energy produced by SNe II and stellar winds. This guarantees that the SNe II particles
are hot enough not to suffer from rapid cooling that is expected in a dense region such as those where stars form
\citep{kpfj02,bto07,dvs12}. 
This also means that we need to resolve the expansion of the bubble from a single SNe II particle, which requires careful choice of the numerical recipe of SPH. We have demonstrated that the new version of {\tt GCD+} is capable of resolving the expansion of a feedback bubble \citep{kogbc13}. Consequently our typical minimum timestep is $\Delta t=100$-1000 yr, similar to what is used in \citet{sdk08} and \citet{hqm12a}.

\subsubsection{Other Feedback Particles}
\label{ofbp-sec}

We apply a similar algorithm to star particles with different IDs. The IMF dictates that about 30 \% of the mass will be ejected after a cosmic time. Therefore star particles with IDs ranging from 9 to 19 turn back into gas particles as a function of their age, which models the mass-loss from intermediate mass stars and SNe Ia feedback. 

For example, ID 9 particles with solar metallicity represent mass-loss from 7 M$_{\odot}$ to 5.6 M$_{\odot}$ stars. When their age of the particle becomes older than the lifetime of a 7 M$_{\odot}$ star, it becomes a feedback particle.  The particle inherits the original metal abundance and receives additional metals that stars in this mass range produce \citep{vdhg97}, which are then diffused via the metal diffusion scheme applied. Their thermal energy is calculated with the SPH scheme, and the additional pressure from these feedback particles are applied to their neighbour particles. For these particles, we turn on radiative cooling during their mass-loss. Once the particle becomes older than the lifetime of a 5.6 M$_{\odot}$ star, the particle becomes a normal gas particle. 

The same algorithm is applied to the particles whose IDs are between 10 and 19, but they are responsible for different mass ranges of stars. The particles whose ID is high enough to cover the mass range of SNe Ia progenitors receive the energy from SNe Ia depending on their age. Unlike the SNe II particles, the feedback particles that received SNe Ia are allowed to be affected by radiative cooling. Because of this algorithm, the particle mass of stars and gas are always constant, and the mass resolution is kept constant. 
   
  The main free parameters of the new version of {\tt GCD+} include energy per SN, $E_{\rm SN}$, and stellar wind energy per massive star, $L_{\rm SW}$. These parameters control the effect of feedback on star formation, which is the most unknown and possibly most important process in galaxy formation and evolution. In this paper, we present the results of two models with and without the stellar energy feedback, and explore the effect of feedback on the star formation in dIrrs. Our fiducial model, Model FB, adopts $E_{\rm SN}=10^{50}$ erg \citep[e.g.][]{tgjs98}\footnote{Various previous studies applied different values of SN energy. However, we think that the effect of feedback depends more on the implementation. Therefore, we think that this value is simply a free parameter which highly depends on the implementation.}
   and $L_{\rm SW}=10^{37}$ erg~s$^{-1}$ \citep[e.g.][]{wmcsm77,mjs80,bg94,om94}. We also present the results of Model noFB, where $E_{\rm SN}=L_{\rm SW}=0$ are adopted. Although we turn off energy feedback in Model noFB, mass-loss and metal enrichment are included, and we track the chemical evolution also in this model.

\begin{figure}
\centering
\includegraphics[width=\hsize]{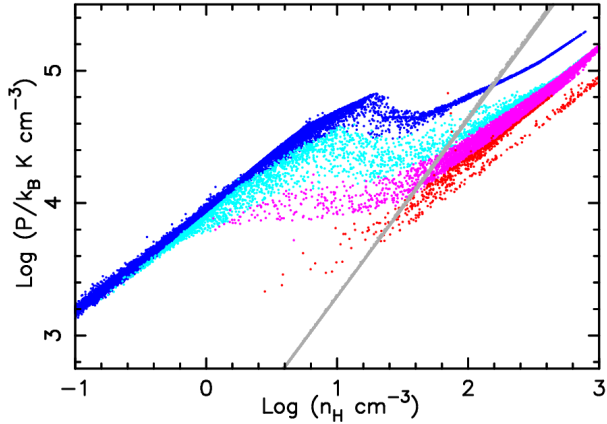}
\caption{
 Thermal pressure, $P/k_{\rm B}$, vs.\ hydrogen number density, $n_{\rm H}$, for gas particles with [Z]$<-0.75$ (blue dots),
  $-0.75<$[Z]$<-0.5$ (cyan dots), $-0.5<$[Z]$<-0.25$ (magenta dots) and [Z]$>-0.25$ (red dots) in Model no FB at $t=1.5$ Gyr.
 where $k_{\rm B}$ is the Boltzmann constant.
The grey line indicates the adopted pressure floor in equation (\ref{peffeq}).
}
\label{pnh-fig}
\end{figure}

\subsection{Dwarf Irregular Simulations}
\label{inic-sec}

 We are interested in investigating the evolution of a dwarf disc galaxy similar in size to dIrr, WLM. We therefore set up an isolated disc galaxy which consists of gas and stellar discs, with no bulge component, in a static dark matter halo potential \citep{rk12,gkc12b}. In this paper, we focus on the effect of feedback on the ISM and star formation in dIrrs. To highlight the evolution of baryonic components and save computational costs, we have fixed the dark matter halo potential. The shape of the dark matter halo could be altered by the evolution of the baryons in small systems like dIrrs, especially if repeated outflows occur owing to strong stellar feedback \citep[e.g.][]{nef96,jrgg05,mcw06,gbm10,pg12,dbmsk13,tpdr13}. However, a live halo component will complicate the evolution of the baryonic disc with effects such as numerical scattering and heating, and may even act as large perturbing masses that artificially disturb the baryonic components if the mass resolution for the dark matter is too low \citep[e.g.][]{dvh13}. It requires a careful set-up and excessive computational resources. Therefore we have elected to model the dark matter halo with static potential. 
 We use the standard Navarro-Frenk-White (NFW) dark matter halo density profile \citep{nfw97}, assuming  a $\Lambda$-dominated cold dark matter ($\Lambda$CDM) cosmological model with cosmological parameters of $\Omega_0=0.266=1-\Omega_{\Lambda}$, $\Omega_{\rm b}=0.044$, and $H_0=71{\rm kms^{-1}Mpc^{-1}}$:

\begin{equation}
\rho _{dm}=\frac{3H_{0}^{2}}{8\pi G}\frac{\Omega _{0}-\Omega_b}{\Omega_0}\frac{\rho _{c}}{cx(1+cx)^{2}},
\end{equation}
where
\begin{equation}
c=\frac{r_{200}}{r_{s}}, \;\; x=\frac{r}{r_{200}},
\end{equation}
and
\begin{equation}
r_{200}=1.63\times 10^{-2}\left(\frac{M_{200}}{h^{-1}M_{\odot }}\right)^{\frac{1}{3}} h^{-1}\textup{kpc},
\end{equation}
\noindent where $\rho _{c}$ is the characteristic density of the profile, $r$ is the distance from the centre of the halo and $r_{s}$ is the scale radius. The total halo mass is set to be $M_{200}=2.0\times 10^{10}M_{\odot }$ and the concentration parameter is set at $c=12$. The halo mass is roughly consistent with the recent measured mass of WLM by \citet{lvbbc12}. However, it is difficult to measure the total mass up to the virial radius, and the total halo mass of the dIrr is not well constrained.

The stellar disc is assumed to follow an exponential surface density profile:
\begin{equation}
\rho _{d,*}=\frac{M_{d,*}}{4\pi z_{d,*}R_{d,*}^2}\textup{sech}^{2}\left(\frac{z}{z_{d,*}}\right)\exp \left(-{\frac{R}{R_{d}}}\right),
\end{equation}
\noindent where $M _{d,*}$ is the stellar disc mass, $R_{d,*}$ is the scale length and $z_{d,*}$ is the scale height. Following the observational estimates of WLM \citep{jsgpw07,kwhn07,lvbbc12}, we adopt $M _{d,*}=1.5\times10^{7}$ M$_{\odot}$, $R_{d,*}=1.0$ kpc and $z_{d,*}=0.7$ kpc. 

The gaseous disc is set up following the method described in \citet{sdh05b}.
The radial surface density profile is assumed to follow an exponential law like the stellar disc with a scale length, $R_{d,gas}=1.5$ kpc. The initial vertical distribution of the gas is iteratively calculated to reach hydrostatic equilibrium assuming the equation of state calculated from our assumed cooling and heating function. The total gas mass is $3.0\times 10^{8}$ M$_{\odot }$. This is significantly higher than the estimated HI gas mass in WLM \citep{kwhn07}. However, it is not straightforward to compare the total gas mass in the simulations with the HI gas mass in observations, because the gas only in the central region can be high enough density to be identified as HI gas, due to UV background radiation. For example, although we have initially set up an exponential profile of the gas disc, the significant amount of the gas in the outer region is quickly evaporated by UV background radiation \citep{ge92,js04}.

 We impose initial radial abundance profiles to both gas and stellar components. The iron abundance profiles for the gas and stars are given by
 \begin{equation}
 {\rm [Fe/H]}(R)=-1.24-0.04\frac{R}{1\ ({\rm kpc})}.
\label{fehreq}
\end{equation}
\citet{lvbbc13} measured the mean metallicity of [Fe/H]$=-1.28\pm0.02$ and the metallicity gradient $d{\rm [Fe/H]}/d(r/r_{c})=-0.04\pm0.04$ dex for WLM. The core radius of WLM is $r_{c}\sim0.5$ kpc \citep{mm98}. Our assumed metallicity gradient is within the error of their measurement. We also add a Gaussian scatter with a dispersion of 0.03 dex and 0.3 dex for gas and stars respectively to the relation of equation (\ref{fehreq}). The $\alpha$-element abundance is then determined by ${\rm [\alpha/Fe]}=0.2$ \citep{cb11} with a Gaussian scatter with a dispersion of 0.02 dex. For the initial star particles, we randomly assign them an age, assuming the age-metallicity relation ${\rm [Fe/H]}=-0.01\times {\rm age(Gyr)}$ and a constant SFR for 10 Gyr.

 We use 3,000,000 gas particles and 150,000 star particles for the initial condition. This leads to 100 M$_{\odot}$ for each particle, and means that both the star and gas particles have the same mass resolution in our simulations. This is not by chance but required from our new modelling of star formation, feedback and metal diffusion shown in Section \ref{code-sec}. We apply the spline softening and variable softening length suggested by \citet{pm07} for SPH particles. We set the minimum softening length at 16 pc (the equivalent Plummer softening length is about 5 pc), which corresponds to the required softening for gas to resolve $n_{\rm H}=10$ cm$^{-3}$ with solar metallicity.  Fig.~\ref{pnh-fig} shows thermal pressure and gas density for gas particles at $t=1.5$ Gyr for Model noFB. The grey line in the figure indicates the adopted pressure floor of equation (\ref{peffeq}). The figure demonstrates that the pressure of some gas particles at $n_{\rm H}>10$ cm$^{-3}$ becomes lower than the pressure floor, i.e. they hit the resolution limit. We therefore used the above softening limit. For the star particles, we applied a fixed spline softening of 16 pc softening length. We run simulations for 1.5 Gyr from the initial conditions described above for Models FB and noFB (Section \ref{ofbp-sec}). We do not include any continuous inflow of the gas for simplicity of numerical setup. Therefore, we do not think that our simulations are capable of studying a long time evolution, which is a reason for stopping the simulation at this short period. This also helps to reduce the total computational cost. Note that in this paper we demonstrate a dramatic evolution even in this short period. We think that our idealised numerical set-up and the short period of evolution for this simulation are a compromised choice and valid for the purpose of this paper.

\begin{figure}
\centering
\includegraphics[width=\hsize]{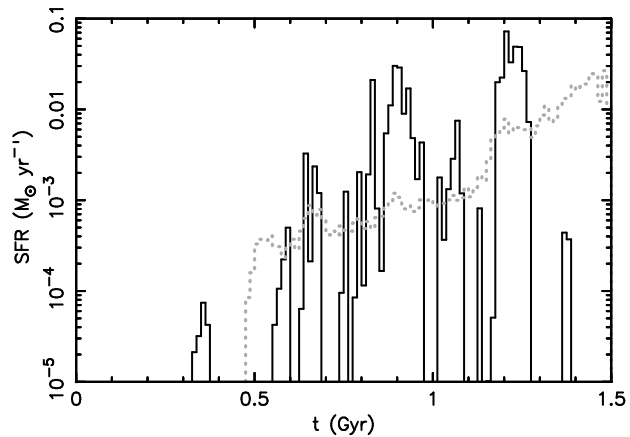}
\caption{
History of SFR of Models FB (black solid) and noFB (grey dotted).
}
\label{sfr-fig}
\end{figure}

\begin{figure*}
\centering
\includegraphics[width=\hsize]{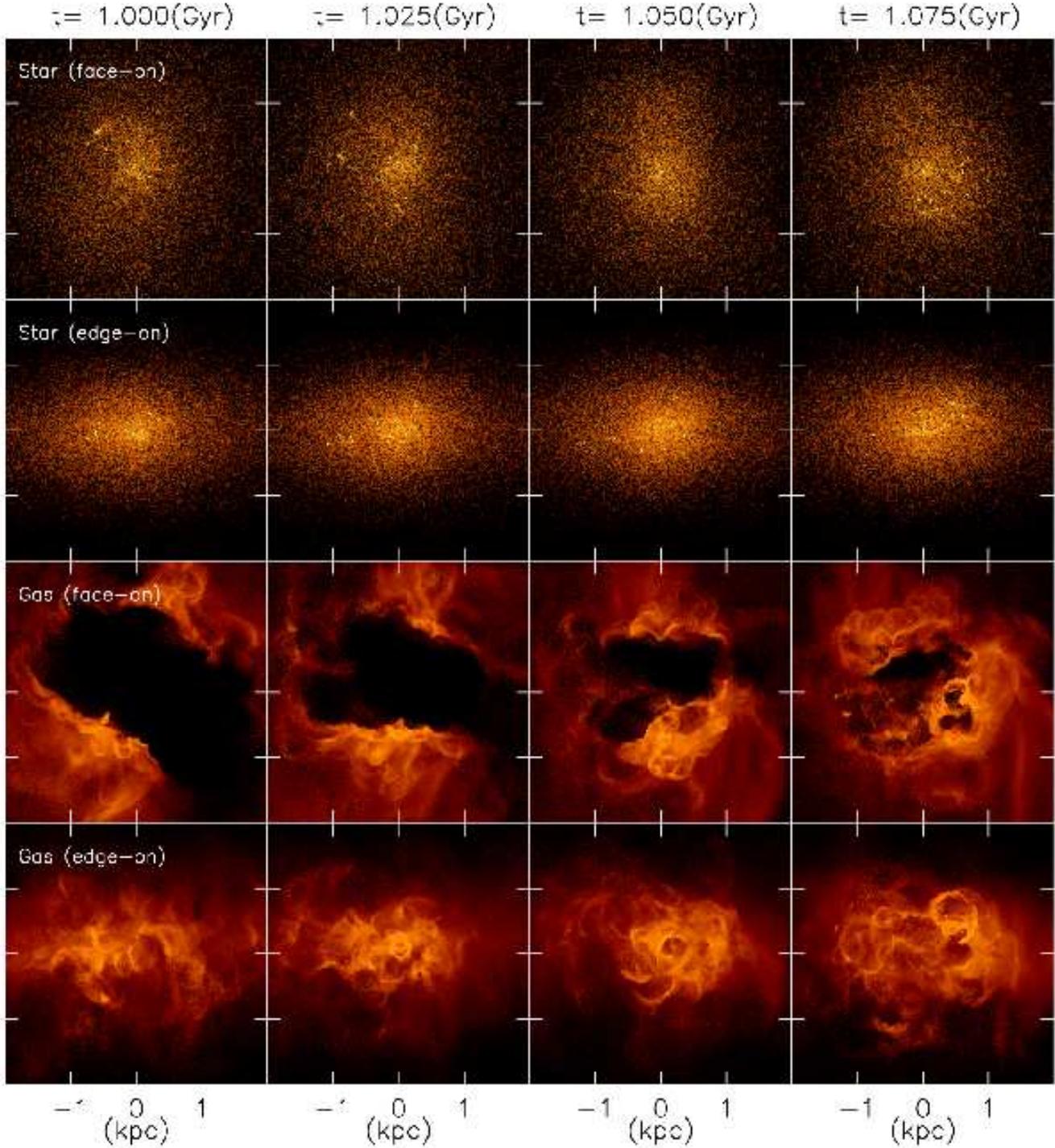}
\caption{
 Snapshots of Model FB from $t=1.0$ Gyr and $t=1.075$ Gyr.
}
\label{evol1-fig}
\end{figure*}

\begin{figure*}
\centering
\includegraphics[width=\hsize]{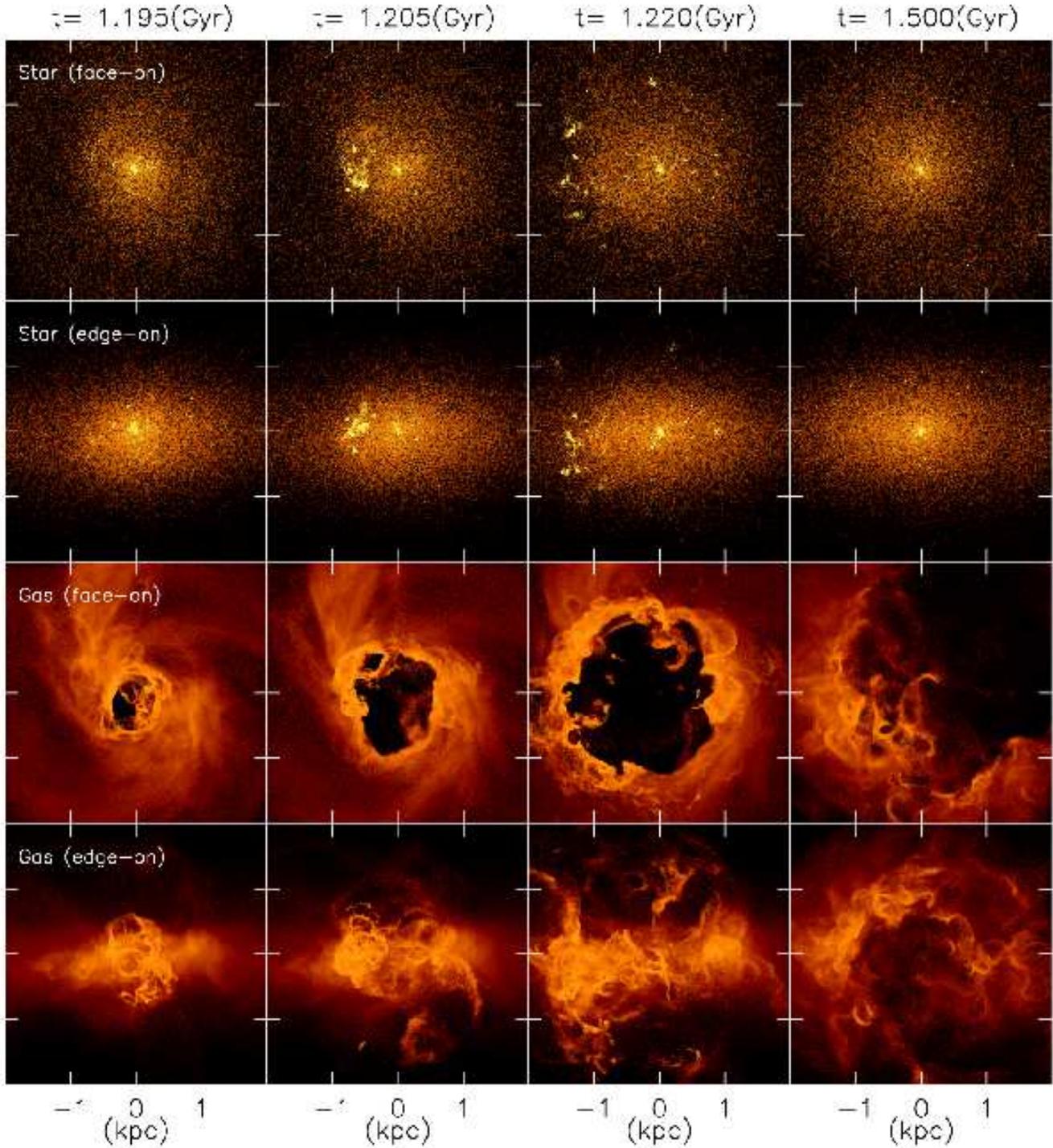}
\caption{
 Same as Fig.~\ref{evol1-fig}, but for the outburst around $t=1.2$ Gyr.
}
\label{evol2-fig}
\end{figure*}

\begin{figure*}
\centering
\includegraphics[width=\hsize]{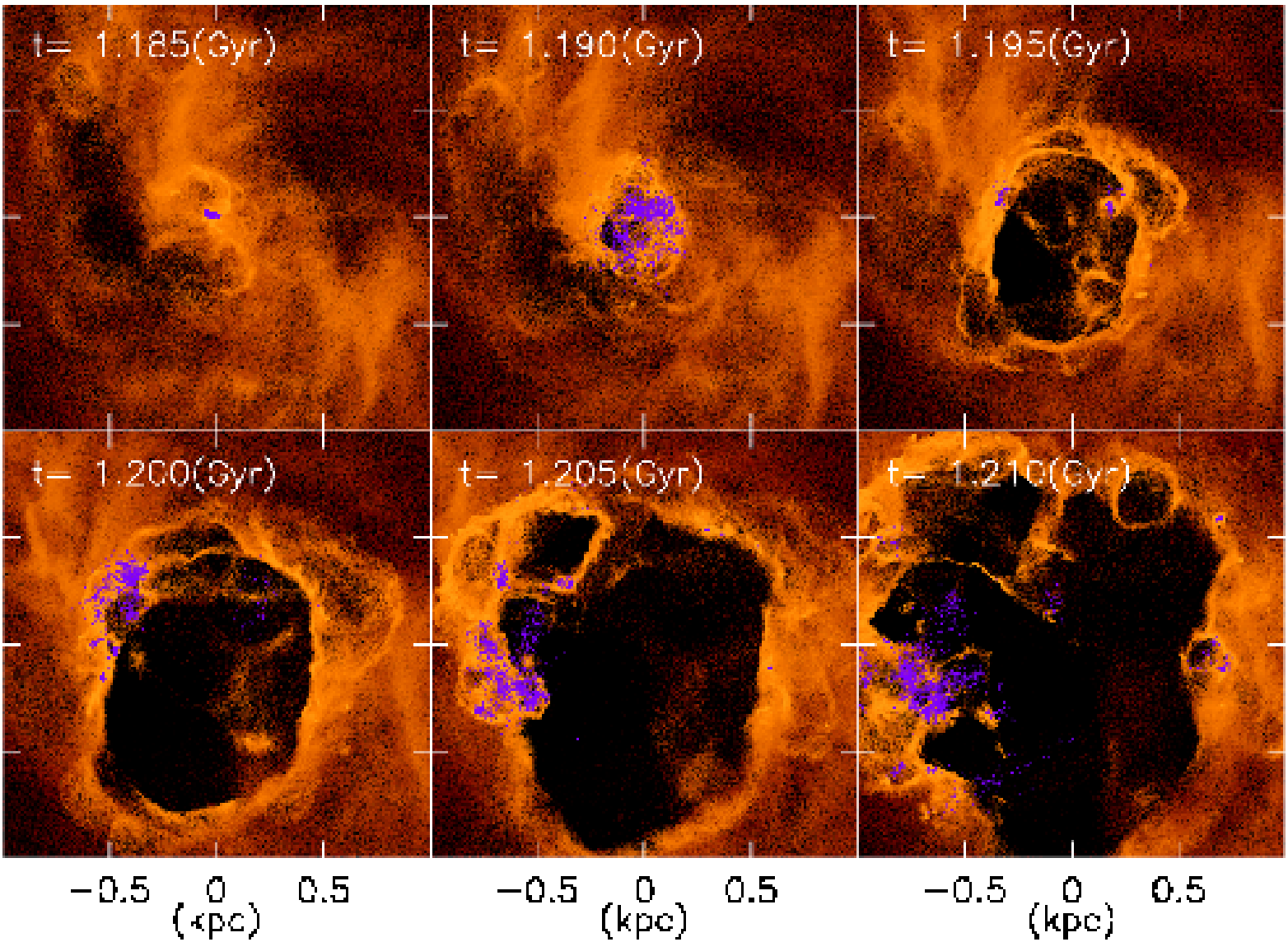}
\caption{
 Close-up face-on view of the evolution of the gas distribution of Model FB. Blue dots indicate the star particles formed within the last 5 Myr.
 }
\label{bubsf-fig}
\end{figure*}

\section{Results}
\label{res-sec}

Fig.~\ref{sfr-fig} shows the history of SFR for Models FB and noFB. The initial gas disc is smooth, and no gas particle has a high enough density to form stars. The gas density gradually increases especially at the centre. At around $t=0.5$ Gyr, star formation starts happening. Fig.~\ref{sfr-fig} clearly shows that the star formation history is quite different between Models FB and noFB. Model FB has stochastic star formation. This is because the gas is blown out, as shown later. The mean SFR from 0.5 Gyr to 1.5 Gyr is about $5\times10^{-3}$ M$_{\odot}$ yr$^{-1}$ for Model FB, which is higher than the estimated mean SFR of WLM, about $2\times10^{-4}$ M$_{\odot}$ yr$^{-1}$ \citep{ad00}, although it is comparable to the current SFR of $6\times10^{-3}$ M$_{\odot}$ yr$^{-1}$ \citep{hel10}.
Note that this is a significantly lower SFR than the recent simulations, including \citet{hqm12a} and \citet{tpdr13}. It could be possible to tune our star formation and feedback parameters to lower the SFR to match the observed mean SFR of WLM. However, the star formation history also depends on the initial condition, and it is difficult to explore the best parameter set from the idealised simulation. It is not our aim to fine-tune the parameters and initial condition for WLM. Rather, we compare two models with and without feedback in identical conditions, and discuss the effect of the feedback on the star formation and the evolution of the ISM. Therefore, our Model FB does not mean our best model. Model FB might be an exaggerated model of feedback effect due to the high SFR, which makes our comparison easier as a numerical experimental study. 

 Model noFB, on the other hand, shows a continuous increase of the SFR. Although at $t=1.5$ Gyr, the SFR reaches more than 0.01 M$_{\odot}$ yr$^{-1}$, it is worth noting that in the earlier phase of the star formation (around $t\sim0.5$-1 Gyr), the SFR in Model noFB is sometimes lower than that of Model FB, and the mean SFR of Model noFB from 0.5 Gyr to 1.5 Gyr is about
 $4\times10^{-3}$ M$_{\odot}$ which is slightly lower than that of Model FB.
  Because of the shallow gravitational potential and inefficient radiative cooling, it takes a long time for enough gas to fall into the central region to maintain the high SFR, if there is no perturbation in the ISM, and the gas smoothly accretes into the central region, like Model noFB. On the other hand, as we demonstrate below, in Model FB, once the star formation begins in the central region, strong feedback produces significant perturbations in the ISM and induces star formation, "the bubble-induced star formation". Note that although the strong stellar feedback can enhance the star formation, i.e. positive feedback \citep[e.g.][]{dbcb05}, 
 for a longer time-scale the overall effect of the stellar feedback reduces the SFR \citep[e.g.][]{jeb13}.
 
Figs.~\ref{evol1-fig} and \ref{evol2-fig} show snapshots of Model FB from $t=1$ Gyr to $t=1.5$ Gyr. After the major burst around $t=0.85$ Gyr, the star formation temporarily completely stops. Fig.~\ref{evol1-fig} shows a hole in the gas whose radius is around 1 kpc at $t=1$ Gyr, which causes the cessation of the star formation. Then, the gas starts re-accreting. Thanks to the inhomogeneous distribution of the gas created by the past generations of the stellar feedback, even during this accretion, the gas collapses locally often in filamentary structures, and forms stars, which corresponds to a low level of SFR in $t=1$-1.1 Gyr in Fig.~\ref{sfr-fig}. Then, once the gas falls into the central region, and the gas density becomes high enough, the concentrated star forming region develops around the centre, and causes massive stellar feedback, which creates bubbles as seen in the snapshot of $t=1.195$ Gyr in Fig.~\ref{evol2-fig}. The expanding bubble collides with the accreting gas. The high-density gas owing to this collision induces the next generation of the star formation. Then, this new generation of star forming regions create the new bubbles in the shell of the original bubbles, which, for example, is seen around $(x,y)=(-0.5, 0.5)$ kpc in the snapshot of the face-on view of the gas at $t=1.205$ Gyr in Fig.~\ref{evol2-fig}. These generations of bubbles and star formation induced by the bubbles propagate outwards, and create a big hole in the gas. This eventually stops star formation again around $t=1.28$ Gyr. Then, the expansion of the bubble stops around $t=1.3$ Gyr, and reaches a radius of more than 2 kpc. Then, the gas re-accretes again, and at $t=1.5$ Gyr it is still in the phase of re-accretion of the gas. This repeated star formation and the cessation due to the strong effect of the stellar feedback is called  "breathing" star formation activity in \citet{sdqkw07} and is demonstrated in the previous simulations \citep[e.g.][]{ccgl01,sdqkw07,tpdr13}.
  
 Our higher resolution simulation enables us to study the bubble-induced star formation in more detail. Fig.~\ref{bubsf-fig} provides a close-up view of the propagation of the star formation owing to the successive generations of the bubbles created by stellar feedback. At $t=1.185$ Gyr, star formation starts at the centre. At this time, there is a plenty of the accreted gas in the central region. Bubbles created by the first generation of star formation collide with the gas in the central region, and create the high-density gas, which leads to the star formation seen in the snapshot at $t=1.19$ Gyr. These new born stars also create bubbles and push the existing bubbles outwards. Fig.~\ref{bubsf-fig} shows that more gas accretes from the upper-left region of these face-on view images. As a result, more star formation occurs in that region at $t=1.2$ Gyr. A bubble with a radius of about 200 pc seen around $(x,y)=(-0.5, 0.5)$ kpc at $t=1.205$ Gyr, was likely created by the young stars highlighted at the $t=1.2$ Gyr snapshot. We can also see from these snapshots that star formation occurs where the smaller bubbles collide with each other, which creates the high-density wall of gas. These snapshots clearly indicate that the successive star formation induced by the collision between the bubbles and/or between the bubbles and the accreting gas are responsible for the continuous expansion of the largest bubble.


\begin{figure}
\centering
\includegraphics[width=\hsize]{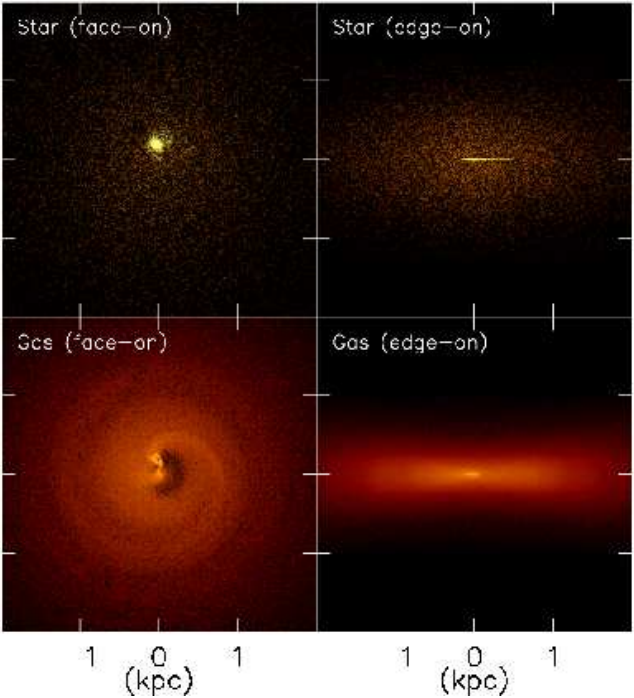}
\caption{
 A snapshot of Model noFB at $t=1.5$ Gyr. 
}
\label{nfd-fig}
\end{figure}

As a comparison, Fig.~\ref{nfd-fig} shows a snapshot of Model noFB at $t=1.5$ Gyr. The figure demonstrates that in this model the gas density can become high enough for star formation only in the central region, and stars are forming only in the central $\sim400$ pc. This is because the gravitational potential is not deep enough to form the large dense disc owing to inefficient radiative cooling. Therefore, the galaxy cannot develop the high-density structures, like spiral arms. The face-on view of the gas distribution in Fig.~\ref{nfd-fig} displays spiral like structure. However, they cannot become dense-enough, because of shallow gravitational potential and inefficient cooling.  Comparison between Models FB and noFB demonstrates that the bubble-induced star formation in Model FB ensures that the star formation spreads widely in the galaxy.

\begin{figure}
\centering
\includegraphics[width=\hsize]{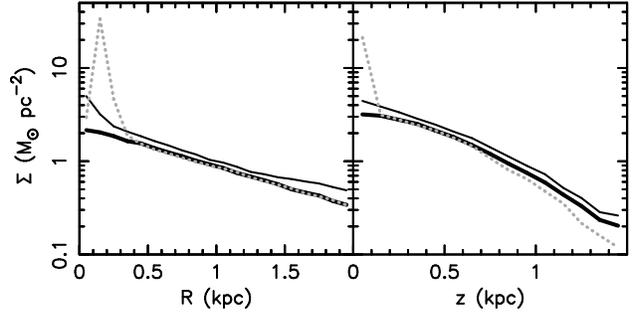}
\caption{
Radial (left) and vertical (right) stellar surface mass density profile for the initial disc (thick black solid) and Motels FB (black solid) and noFB (grey dotted) at $t=1.5$ Gyr. The vertical surface mass density was measured within the range of $0<x<0.5$ kpc. 
}
\label{mpr-fig}
\end{figure}

 Our results infer that there could be a critical total mass or baryonic mass of the galaxy which separates the galaxies in which the bubble-induced star formation is dominant from the galaxies which harbour a gas disc with high enough density to develop local instabilities and non-axisymmetric structures, such as spiral arms, that may trigger star formation. Unfortunately, using our idealised simulations, we cannot explore the critical mass which must be sensitive to many different factors, such as the shape of dark matter halo and stellar disc, mass accretion history, background radiation field. Also, exploring the critical mass is not the aim of this study. 

 
 Rather, our idealised simulations show that there can be small enough galaxies where spiral arms cannot be developed due to too shallow potentials and inefficient cooling. In such small galaxies the bubble-induced star formation can make the star formation area bigger. In this case, the size of the galaxy could be determined where the density of the ISM becomes so small that the expanding bubbles no longer accumulate enough ISM to build up high-density shells. Edge-on views in Fig.~\ref{evol2-fig} show that this is true also for the perpendicular direction to the disc. The bubbles and star forming regions also propagate perpendicular to the disc. This can explain a high scale height of dIrrs. Fig.~\ref{mpr-fig} demonstrates this more quantitatively. Fig.~\ref{mpr-fig} shows the radial and vertical stellar surface density profiles for Models FB and noFB at $t=1.5$ Gyr, comparing with the initial  surface density profiles. In Model noFB, the stellar mass increases only in the central region ($R<400$ pc)\footnote{Radial surface density profile of Model noFB shows an offset between the centre of the dark matter halo and stellar density peak. This is also seen in Fig.~\ref{nfd-fig}. Note that the centre of Figs.~\ref{evol1-fig}-\ref{mpr-fig} is set to be the centre of the dark matter halo.} On the other hand, although there is a small excess in the central region, Model FB shows the increase of the stellar surface density at a wide range of radii and vertical height, and the overall shape of both radial and vertical profiles at $t=1.5$ Gyr is similar to the initial profile.

 
Fig.~\ref{evol2-fig} also shows that there are some stars escaping from the disc region. These could be identified as halo stars, as discussed in the previous simulation studies \citep{sdqgk09}. It is also worth noting that because the ISM is continually disturbed by many generations of stellar feedback bubbles, the re-accretion of the gas is not circularly symmetric, and the star forming region owing to the collisions with bubbles are not symmetric, e.g. a snapshot of $t=1.205$ Gyr in Fig.~\ref{bubsf-fig}. Such non-symmetric star forming regions and bubbles are often observed in dIrrs \citep[e.g.][]{kwhn07}, which can be also explained by the bubble-induced star formation demonstrated by Model FB.

\begin{figure}
\centering
\includegraphics[width=\hsize]{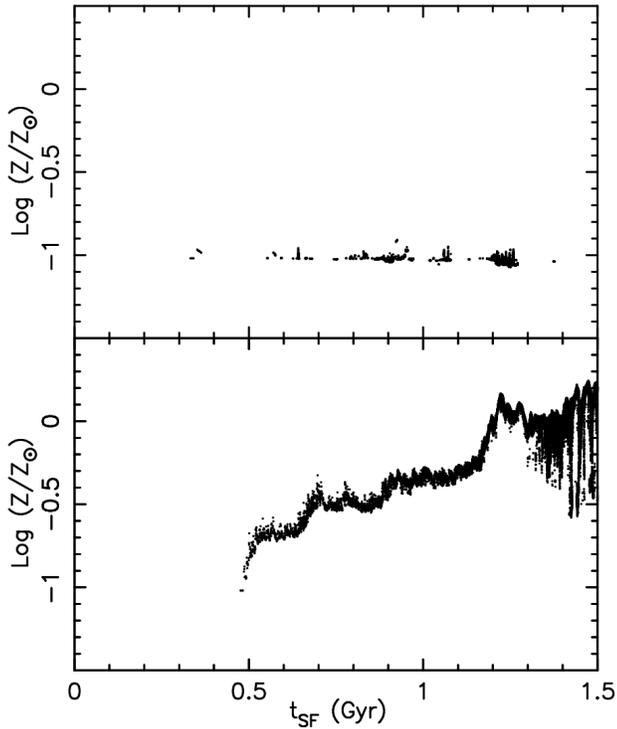}
\caption{
 Metallicity of new born stars as a function of the formation time for Models FB (upper) and noFB (lower). 
}
\label{Ztsf-fig}
\end{figure}

\begin{figure}
\centering
\includegraphics[width=\hsize]{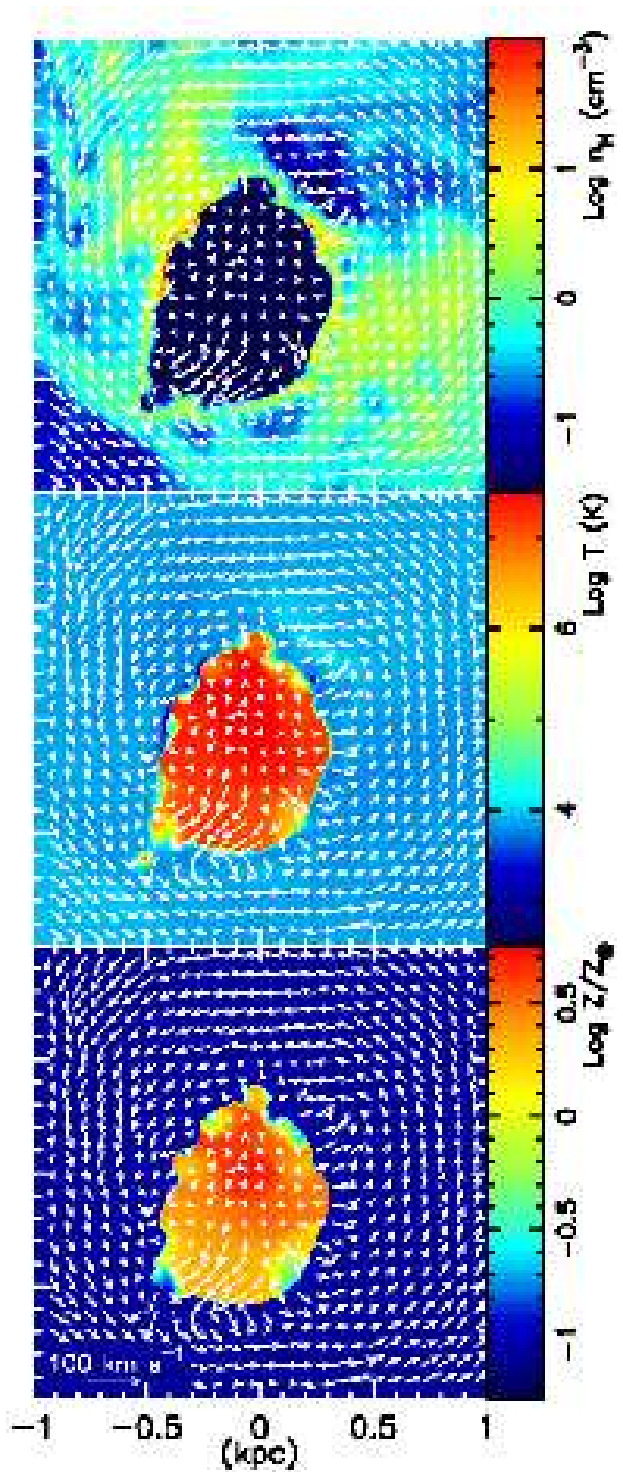}
\caption{
  Hydrogen number density (top), temperature (middle) and metallicity (bottom) of the gas at the disc plane, $z=0$, of Model FB at $t=1.195$. The arrows (same in all the panels) indicate the gas velocity field.
}
\label{gslice-fig}
\end{figure}

\subsection{Metallicity}
\label{metal-sec}

Fig.~\ref{Ztsf-fig} shows the metallicity of the stars formed during the simulations. Model FB shows that the metallicity of the new born stars is almost constant, and kept low. On the other hand, in Model noFB the metallicity of the new born stars increases with their formation time. In this model stars are forming only in the central region. The new born stars enrich the gas only in the central region. The metallicity of the gas in the central region quickly goes up. Therefore, the later generation of stars inevitably become metal-rich. The metallicity reaches higher than the solar metallicity. The mean metallicity of the stars formed within the last 100 Myr is ${\rm [Z]}=0.023$, and this is unacceptably high. In Model FB, a similar amount of star formation took place. However, the metallicity of the new born stars is low. This is because the galaxy has a large reservoir of the metal-poor ISM and the bubble-induced star formation enhances the mixing between the ISM and the metal-rich bubbles.  Fig.~\ref{gslice-fig} displays hydrogen number density, temperature, metallicity and velocity field of the gas at the disc plane of Model FB at $t=1.195$ Gyr. As expected, the metallicity is very high within the bubble. However, the density within the bubble is too low and temperature is too high for star formation.  As demonstrated above, the star formation can happen only within the high-density shells of the bubbles. Fig.~\ref{gslice-fig} demonstrates that the cold high-density shells have metallicity as low as the surrounding ISM, which indicates that the ejected metals from feedback particles are well-mixed with the ISM, when the gas density reaches the star formation threshold. In other words, the bubble-induced star formation guarantees the widely spread star formation, and also keeps the metallicity of the star forming region low. Note that the dispersion of the metallicity in Model FB is too small, and it is inconsistent with the variation of the estimated abundances for the observed young stars in WLM which shows [Z] from about $-0.5$ dex to $-1$ dex \citep{ukbpg08}. We therefore think that the metal diffusion in our simulations is likely to be overestimated. We need to calibrate the metal diffusion model more. Comparisons between more observations in various dIrrs and the simulations like those used in this paper would be an effective way to improve the metal diffusion model. Calibrating the efficiency of metal diffusion via metallicity distribution function echoes the conclusions drawn by \citet{pgbcs12} and \citet{gpbsb13}.

\subsection{Kinematics}

It is interesting to see from the velocity field of Fig.~\ref{gslice-fig} that some bubble shells are expanding faster than the rotation speed of the ISM, and some shells are moving against the rotational direction of the galactic disc. It can also be expected that if the bubbles are expanding from stars that are more slowly rotating than the gas, they collide with the ISM more violently on the back side, i.e. the opposite side to the rotation direction, than on the front side. Therefore, if the bubble-induced star formation is the dominant mechanism of star formation in dIrrs, we expect that more young stars are rotating slower than the ISM.  Because the rotation velocity of dIrrs is small and can be smaller than the expansion speed of some bubbles, we expect many counter-rotating stars. 

\begin{figure}
\centering
\includegraphics[width=\hsize]{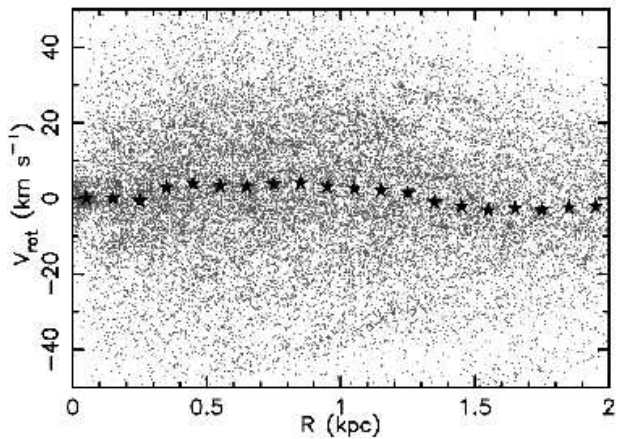}
\caption{
 The rotation velocity as a function of the radius of the stars born (grey dots) during the 1.5 Gyr evolution for Model FB at $t=1.5$ Gyr. The black stars indicates the mean rotation velocity at the different radii. 
}
\label{vr-fig}
\end{figure}

Fig.~\ref{vr-fig} displays the rotation velocity of the stars formed in the simulation period, i.e. $t_{\rm SF}=0-1.5$ Gyr,  as a function of radius for Model FB at $t=1.5$ Gyr. The figure shows that there are a significant number of counter-rotating stars, and the mean rotation speed is almost zero. Fig.~10 of \citet{lvbbc12} shows that in WLM especially within 1 kpc (assuming 1 arcmin$=0.25$ kpc) the mean rotation velocity of stars are as low as 5 km s$^{-1}$. This is similar to the maximum rotation in Model FB, $V_{\rm rot,max}\sim4$ km s$^{-1}$. We note that our idealised simulations do not include the later gas accretion from the inter-galactic medium, and our simulation likely underestimates the rotation of the ISM especially in the outer region. Feedback of Model FB might be too high, too. Nevertheless, we still think from these results that the significant number of counter-rotating stars are an inevitable consequence of the bubble-induced star formation. It would be an interesting observational test to measure the mean rotation velocity and the fraction of the counter-rotating stars more accurately.

\section{Summary}
\label{sum-sec}

 We present numerical simulations of the evolution of a dIrr similar in size to WLM using the updated version of our original N-body/SPH chemodynamics code, {\tt GCD+}. Our high-resolution simulations enable us to study how the strong feedback affects the ISM and star formation in small galaxies.
 
 The simulation without stellar energy feedback has star formation only in the central region, where the gas density can become high enough for star formation. In this small system, the gravitational potential is shallow and radiative cooling is inefficient. As a result, the gas disc is stable to local instabilities except in the central region, and consequently no structures such as spiral arms can develop. Therefore, the smooth gas accretion to the central region is the only way to reach a high enough gas density to form stars. 
 
  On the other hand, our strong feedback model demonstrates that once the bubbles are created by stellar feedback, the bubbles can collide with the ISM and/or the other bubbles, and create the dense filaments, where the gas density becomes high enough to form the next generation of stars. Such generations of stars can propagate outwards until the ISM density becomes too low for the bubbles to accumulate enough ISM to build up the dense shells. These successive bubbles induced by the generations of star formation can therefore produce the stars in the larger area compared to the model without the stellar energy feedback. 
  
   Our simulation with feedback also demonstrates that despite the large amount of metals produced by the new generation of stars, the metallicity of the star forming region is kept low, because the metals produced are well mixed with the metal-poor ISM before they reach high enough density for star formation. We also find that the bubble-induced star formation leads to significantly lower rotation velocity of new born stars, compared to the gas. As a result, in a small system like dIrr, there will be many counter-rotating stars produced if the bubble-induced star formation is a dominant mechanism in dIrr.
   
   We conclude that the bubble-induced star formation is one way to maintain a spread star formation, low metallicity, low stellar rotation velocity and high stellar velocity dispersion observed in dIrr. This is different from spiral arm induced star formation \citep{wr69,be11}, where the star formation predominantly occurs in and around the spiral arms as seen in larger spiral galaxies both in observational studies \citep[e.g.][]{rk90,frdlw11,fckph12}  and in numerical simulations \citep[e.g.][]{wk04,kw08,dp10,wbs11,gkc12b}. Still, the bubble-induced star formation can be important in large spiral galaxies, as seen in the superbubbles in the Milky Way \citep[e.g.][]{owkw05}, although the scale of the influence would be relatively smaller in the large galaxies, because of the higher density in the gas disc. 
Also, significant velocity dispersion of the HI gas is often observed in spiral galaxies \citep{dbb06,trlmw09}, which can be maintained by the bubble-induced star formation as shown in our lower resolution simulations \citep{rk12}. Although it is computationally challenging, we are preparing for the simulations of larger spiral galaxies with similar physical resolution to this paper. We will explore the effect of bubble-induced star formation in larger spiral galaxies, and compare with the case of small galaxies.

 \section*{Acknowledgments}
We thank anonymous referee for their constructive comments and helpful suggestions which have improved the manuscript.
BKG acknowledges the support of the UK¡Çs Science \& Technology Facilities Council (STFC, ST/J001341/1). The calculations for this paper were performed on the UCL Legion, the Cray XT4 at Center for Computational Astrophysics (CfCA) of National Astronomical Observatory of Japan and the DiRAC Facilities (through the COSMOS consortium) jointly funded by STFC and the Large Facilities Capital Fund of BIS.   
We also acknowledge PRACE for awarding us access to resource Cartesius based in Netherlands at SURFsara.
 
\bibliographystyle{mn}

\appendix

\section[]{Lower resolution model}
\label{lrm-sec}

\begin{figure}
\centering
\includegraphics[width=\hsize]{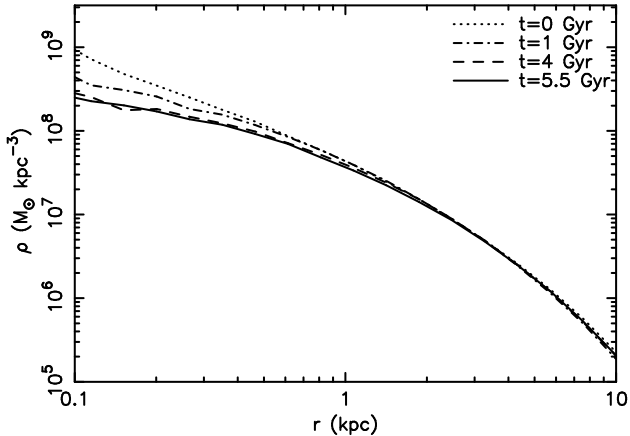}
\caption{
Radial dark matter mass density profile for Model LRLH at $t=0$ Gyr (dotted), $t=1$ Gyr (dot-dashed),
 $t=4$ Gyr (dashed) and $t=5.5$ Gyr (solid). 
}
\label{dmprlr-fig}
\end{figure}

\begin{figure}
\centering
\includegraphics[width=\hsize]{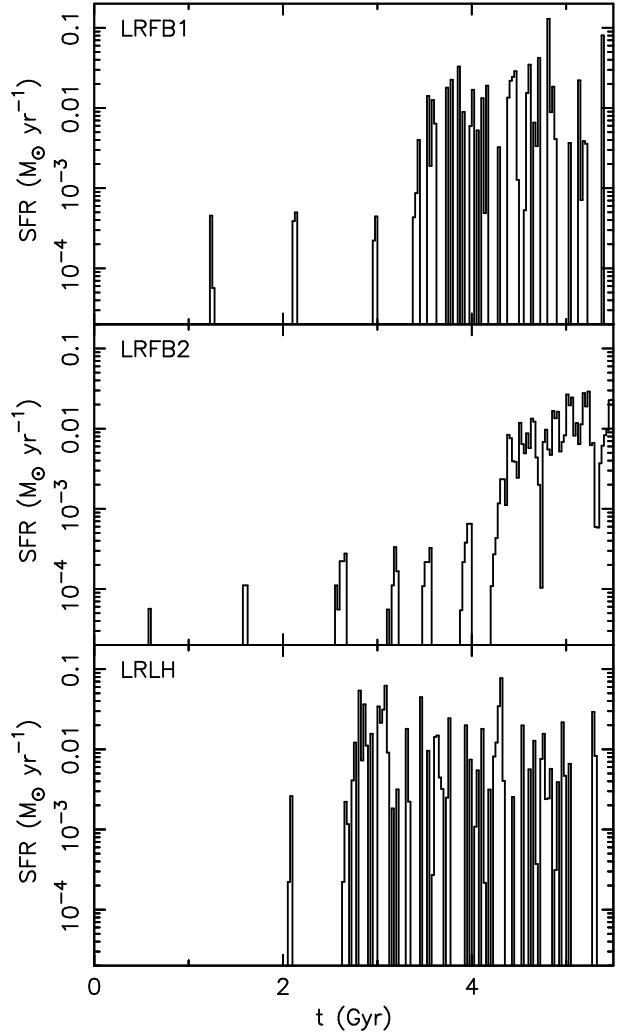}
\caption{
History of SFR of Models LRFB1 (upper), LRFB2 (middle) and LRLH (bottom).
}
\label{sfrlr-fig}
\end{figure}

\begin{figure}
\centering
\includegraphics[width=\hsize]{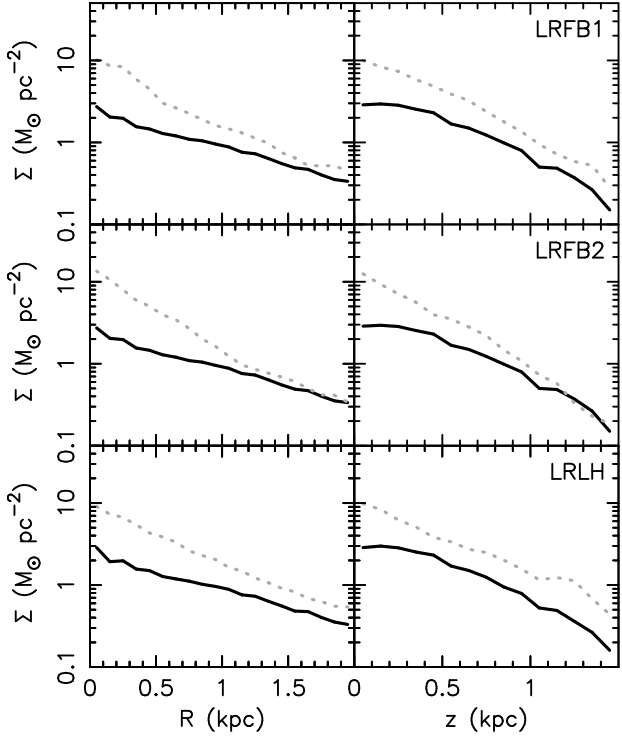}
\caption{
Radial (left) and vertical (right) stellar surface mass density profile for the initial disc (thick black solid) and the disc
at $t=5.5$ Gyr (grey dotted) for Models LRFB1 (upper), LRFB2 (middle) and LRLH (bottom).
The vertical surface mass density was measured within the range of $0<x<0.5$ kpc. 
}
\label{mprlr-fig}
\end{figure}

\begin{figure}
\centering
\includegraphics[width=\hsize]{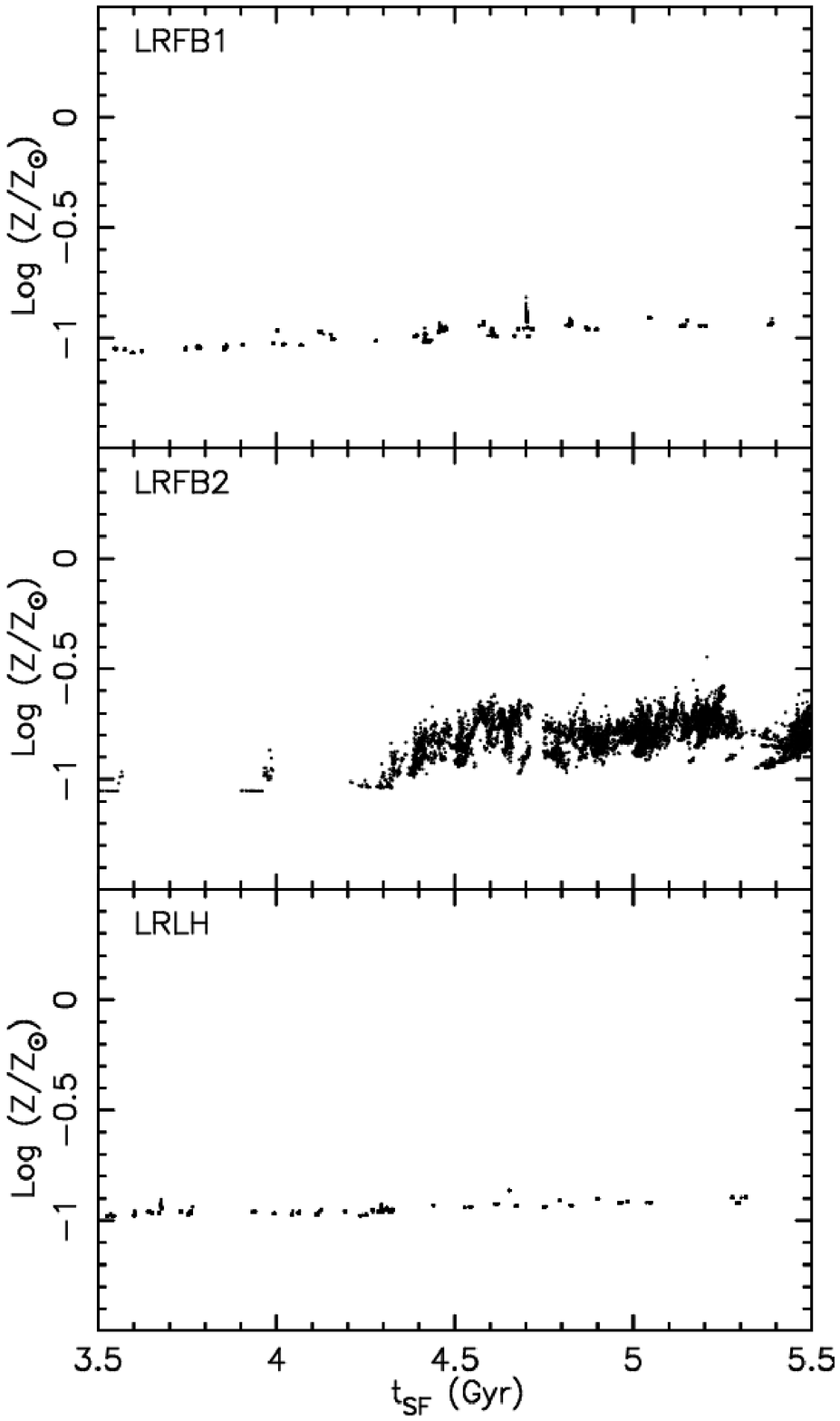}
\caption{
 Metallicity of new born stars as a function of the formation time for Models LRFB1 (upper), LRFB2 (middle)
 and LRHR (bottom).
 There appear to be a fewer particles in Models LRFB1 and LRHR,
 because there are many star particles with almost same metallicity and formation time.
}
\label{Ztsflr-fig}
\end{figure}

We present the results of three complementary lower-resolution models, Models LRFB1, LRFB2 and LRLH. All models use ten times fewer particles, i.e. the mass of star and gas particles is 1,000 M$_{\odot}$. We employ the minimum softening length of 34 pc
for the baryon particles. 
Note that our stellar feedback scheme is not designed to be resolution independent. Instead, we adopt a simple scheme, which has less parameters, but is sensitive to the resolution. We can still calibrate the feedback parameters at a fixed physical resolution. The aims of these complementary models are to demonstrate that our qualitative conclusion is robust against the numerical resolution, 
 the choice of star formation parameters, such as $n_{\rm H,th}$ and $C_*$ (see Section \ref{code-sec}), and if or not
a live dark matter halo is adopted.

 Model LRFB1 is similar to Model FB, and uses $n_{\rm H,th}=1000$ cm$^{-3}$ and $C_*=1.0$. On the other hand, Model LRFB2 employs $n_{\rm H,th}=10$ cm$^{-3}$ and $C_*=0.02$. Fig.~\ref{sfrlr-fig} shows the SFR history for Models LRFB1 and LRFB2.
Model LRLH employs the same $C_*$ and $n_{\rm H,th}$ as Model LRFB1. However, Model LRLH adopts a live dark matter halo,
which is described by $10^{4}$ M$_{\odot}$ collisionless particle with the softening length of 73 pc. 
To reduce the total number of dark matter particles, we adopted a truncated NFW profile \citep{ras09},
\begin{equation}
\rho _{dm}=\frac{3H_{0}^{2}}{8\pi G}\frac{\Omega _{0}-\Omega_b}{\Omega_0}\frac{\rho _{c}}{cx(1+cx)^{2}}
 \exp(-x^2 r_{200}^2/r_t^2),
\end{equation}
where $r_t$ is the truncation radius, and we set $r_t=13.9$ kpc.
We apply $c=20$ for the initial condition of Model LRLH which is significantly higher than $c=12$ used for the fixed halo case.
However, we find that the dark matter density profile becomes shallower in the live halo case (Fig.~\ref{dmprlr-fig}), which leads to
a similar profile to the NFW profile with $c=12$.
 
  Comparing Model LRFB1 with Model FB (Fig.~\ref{sfr-fig}), although Model LRFB1 shows stochastic star formation due to the bubble-induced star formation at the later epoch, it starts much later ($t\sim3.5$ Gyr) than Model FB. This is because in the lower-resolution model it takes a longer time to accumulate enough gas into the central region to cause the bubble-induced star formation and outward propagation of star forming region. In addition, at the earlier time the central region is fragile against stellar feedback, because less gas accretes into the central region, even when the gas density reaches high enough density for star formation at the very centre. This causes the three bursts (t$\sim 1.2, 2.1$ and 3.0 Gyr) with a longer quiescent period in-between. 
  
 Because of the lower threshold density for star formation, Model LRFB2 has the first burst at an earlier epoch ($t\sim0.6$ Gyr). 
 However, at this time the central region does not have enough gas accumulated to cause the bubble-induced star formation,
 and is fragile against stellar feedback. This delays the accumulation of gas in the central region, and the onset of the bubble-induced
 star formation. The SFR history around $t\sim4$-5.5 Gyr for Model LRFB2 looks smoother and less bursty than that for Model LRFB1. However, we observe the bubble-induced star formation, i.e. the star formation happens in the shells of the bubbles, and propagates outward.
 
Because of the higher concentration, which helps accumulating enough gas into the central region, Model LRLH starts 
the bubble-induced star formation earlier ($t\sim2.6$ Gyr) than the other models.
It is surprising to see that the first burst of star formation is later ($t\sim2.0$ Gyr) in Model LRLH than Model LRFB1 ($t\sim1.2$ Gyr).
This may be because the central density of the dark matter halo is limited at the scale of the resolution in the live dark matter halo,
while the fixed dark matter halo follows the NFW profile without any limit. 
 
  The mean SFR for the last 1 Gyr ($t=4.5$-5.5 Gyr) is $9.5\times10^{-3}$ M$_{\odot}$ yr$^{-1}$ for Model LRFB1 and $1.1\times10^{-2}$ M$_{\odot}$ yr$^{-1}$ for Model LRFB2. These SFR is slightly higher than that for Model FB. As we mentioned above, our feedback scheme is resolution dependent, and it is not surprising to see the difference between models with different resolutions. Models LRFB1 and LRFB2 show similar mean SFR. This demonstrates that both $n_{\rm H,th}$ and $C_*$ affect the resultant SFR, and applying high $n_{\rm H,th}$ and $C_*$ has a similar effect to using low $n_{\rm H,th}$ and $C_*$. 
The mean SFR for the last 1 Gyr for Model LRLH is $3.6\times10^{-3}$, which is smaller but similar to that for Model LRFB1.
 
Fig.~\ref{mprlr-fig} shows the radial and vertical surface mass density profile at $t=0$ and 5.5 Gyr for 
Models LRFB1, LRFB2 and LRLH.
 Similar to Model FB in Fig.~\ref{mpr-fig}, all the
 models show the increase in the stellar surface density at a wide range of radii and vertical height. Model LRFB2 shows more increase in the inner region, which indicates more star formation in the inner region.

Fig.~\ref{Ztsflr-fig} shows the metallicity of stars formed in the last 2 Gyr of the evolution for 
Models LRFB1, LRFB2 and LRLH.
Similar to Model FB in Fig.~\ref{Ztsf-fig}, the metallicity of new born stars is kept low, because of the efficient mixing between the metal-rich bubbles and the metal-poor ISM. Because of the high SFR, Model LRFB2 shows higher metallicity. Also, the low density threshold of star formation in Model LRFB2 causes more scatter in the metallicity, because the stars can form from lower-density gas where the metallicity is more inhomogeneous (Fig.~\ref{gslice-fig}). As discussed in Section \ref{metal-sec}, our metal diffusion model needs to be calibrated more, and it is not independent from the star formation parameters. 

Although there are several (expected) differences between the high- and low-resolution models, the overall results from the lower-resolution simulations are qualitatively consistent with the higher-resolution models. This reassures our conclusion about the bubble-induced star formation: the level of star formation and metallicity is kept low, while stars form in a widely spread area. 

\label{lastpage}

\end{document}